# Competitive Dynamics on Complex Networks

Jiuhua Zhao, Qipeng Liu & Xiaofan Wang[*]

Department of Automation, Shanghai Jiao Tong University, and Key Laboratory of System Control and Information Processing, Ministry of Education of China, Shanghai 200240, China.

**We consider a dynamical network model in which two competitors have fixed and different states, and each normal agent adjusts its state according to a distributed consensus protocol. The state of each normal agent converges to a steady value which is a convex combination of the competitors' states, and is independent of the initial states of agents. This implies that the competition result is fully determined by the network structure and positions of competitors in the network. We compute an Influence Matrix (IM) in which each element characterizing the influence of an agent on another agent in the network. We use the IM to predict the bias of each normal agent and thus predict which competitor will win. Furthermore, we compare the IM criterion with seven node centrality measures to predict the winner. We find that the competitor with higher Katz Centrality in an undirected network or higher PageRank in a directed network is much more likely to be the winner. These findings may shed new light on the role of network structure in competition and to what extent could competitors adjust network structure so as to win the competition.**

Competition among a set of competitors for obtaining a maximum number of votes from other agents in a social network is a both important and common phenomena in real world. The competitors could be candidates in numerous leader-selection cases, ranging from head–election in a small group to president-election in a whole country[1]. They could also be those who have different proposals or promote different brands of a product such as mobile phone and car[2]. There have also been some researches on, for example, how the fractions of speakers of several competing languages evolve in time[3] and even how the emerging Bitcoins appear to be a possible competitor to usual currencies[4].

The most well-known model in social dynamics for the competition of species is the voter model[5,6], which has also later on been used for the analysis of diffusion of innovations and consumption decisions. In its simplest form, each agent in the voter model holds one of the two states. At each time step, a randomly selected agent takes the state of one of its neighbors. Over the years, many modifications and extensions of the original voter model have been proposed[7]. Voter-like dynamics on networks with different topologies and the interplay between topology and dynamics have also been investigated[8,9]. However, many of such models, including the voter model, Sznajd model[10,11], Deffuant model[12], Hegselmann-Krause model[13] and so on have been focused on whether full consensus can be reached.

A nature way to consider the existence of competitors in a network is to view them as zealots[14,15] or stubborn agents[16-18] with fixed and different states. For example, it is shown that the existence of competing zealots in the voter model prevents convergence and results in fluctuations in regular lattices[14] and complete graphs[15]. Competitive dynamics with continuous states in the stochastic gossip model is investigated in Ref.16, in which long-run disagreements and persistent fluctuations appear. Influence of network structure and locations of stubborn agents on the fluctuation of final states in a binary opinion formation model is studied in Ref. 17. In Ref. 18，given one set of stubborn agents as mis-informers (agents who spread misinformation), the placement of the other set of stubborn agents (named information disseminators) is formulated as an optimization problem.

The question we address in this work is: *How do positions of competitors in a network affect voting outcome? That is, can we predict which competitor will win* *in the sense that majority of agents in the network will eventually support the competitor? Can we predict which competitor a normal agent will support based on the network structure?* Intuitively, the problem of which competitor will win should be related to the relative impact of the competitors in a network. How to characterize the impact or importance of an individual (or even a community) in a network is a question of great importance and applications in network analysis. Traditionally, identifying such influential nodes usually relies on concepts of centralities, including degree (DC), betweenness (BC)[19], closeness (CC)[20], eigenvector centrality (EC)[21], Katz centrality (KC)[22], PageRank (PR)[23], and so on. Recently, a lot of researches have also been focused on identifying influential nodes in dynamical processes on networks. For example, Kitsak *et al.* have argued that there are circumstances in which a node with the highest DC or the highest BC has little effect, and the most efficient spreaders are those located within the core of the network as identified by the k-shell decomposition[24]. However, till now, we still lack an understanding on which of these measures could best predict the winner among competitors in a network.

**Results**

**A dynamic model for competition.** We consider a directed and weighted network with $N$ agents and $M$ links. The agent set is denoted as $V = \{1, 2, \cdots, N\}$ and the topology of the network is described by a coupling matrix $A = (a_{kl})_{N \times N}$: if agent $k$ is directly influenced by agent $l$, then there is a link from agent $k$ to agent $l$ and $a_{kl} > 0$; otherwise, $a_{kl} = 0$. For simplicity, we assume that there are just two competitors in the network, denoted as agents $i$ and $j$, which have fixed and different states as follows:

$$x_i(t) \equiv +1, \ x_j(t) \equiv -1, \ \forall t \geq 0. \tag{1}$$

Every other agent (called normal agent) $k \in V/\{i,j\}$ has an initial state randomly chosen from $[+1,-1]$ and updates its state as follows:

$$x_k(t+1) = x_k(t) + \varepsilon \sum_{l \in N_k} a_{kl}\left(x_l(t) - x_k(t)\right), \qquad (2)$$

where $x_k(t)$ is the state of agent $k$ at time $t$; the parameter $\varepsilon$ captures the level of neighbors' influence; $N_k = \{l \in V \mid a_{kl} > 0\}$ is the set of neighboring agents of agent $k$ that can directly influence agent $k$. Note that Eq. (2) belongs to a set of distributed consensus protocols, which can be traced back to the classical model of DeGroot[25]. However, the existence of competitors in the network prohibits global consensus. Instead, we have the following convergence result:

Suppose that

1) Each normal agent has a path connecting to at least one competitor;
2) $0 < \varepsilon < D_{\max}^{-1}$, where $D_{\max}$ is the largest out-degree of agents in the network.

Then the state of each normal agent will eventually reach a steady value, i.e., as $t \to \infty$,

$$X_{norm}(t) \to \bar{X} \triangleq (\bar{D} - \bar{A})^{-1}\begin{bmatrix} \mathbf{c}_i & \mathbf{c}_j \end{bmatrix}\begin{bmatrix} +1 \\ -1 \end{bmatrix}, \qquad (3)$$

where $X_{norm} \in R^{N-2}$ represents the state vector of all normal agents, and $\bar{D}$, $\bar{A}$ and $\begin{bmatrix} \mathbf{c}_i & \mathbf{c}_j \end{bmatrix}$ can all be derived from the network coupling matrix $A$. Furthermore, if $x_k(0) \in [-1, +1]$, $\forall k \in V/\{i,j\}$, then $x_k(t) \in [-1, +1]$, $\forall t > 0$. The detailed analysis can be found in Methods.

$\bar{x}_k > 0$ ($\bar{x}_k < 0$) implies that agent $k$ will finally support competitor $i$ ($j$), and $|\bar{x}_k|$ corresponds to the degree of supporting. $\bar{x}_k = 0$ implies that agent $k$ will be a neutral agent which does not support any competitor. Denote

$$\Phi_{ij} \triangleq \sum_{k \in V/\{i,j\}} \text{sgn}(\bar{x}_k), \qquad (4)$$

where $\text{sgn}()$ is the sign function. If $\Phi_{ij} > 0$, then competitor $i$ will win in the sense that more normal agents will support him; if $\Phi_{ij} < 0$, competitor $j$ will win; if $\Phi_{ij} = 0$, the competition ends up with a draw.

**An illustration example.** Fig. 1 shows the competitive dynamics on three simple undirected networks which have the same number of agents but different coupling structures. We take agent 1 and agent 10 as two competitors in each network with fixed states $x_1 \equiv +1$ and $x_{10} \equiv -1$. Steady states of normal agents are computed according to Eq. (3). An red (blue) node represents an agent with positive (negative) state. The darker the color the larger the absolute value of the state. Nodes with white color represent neutral agents.

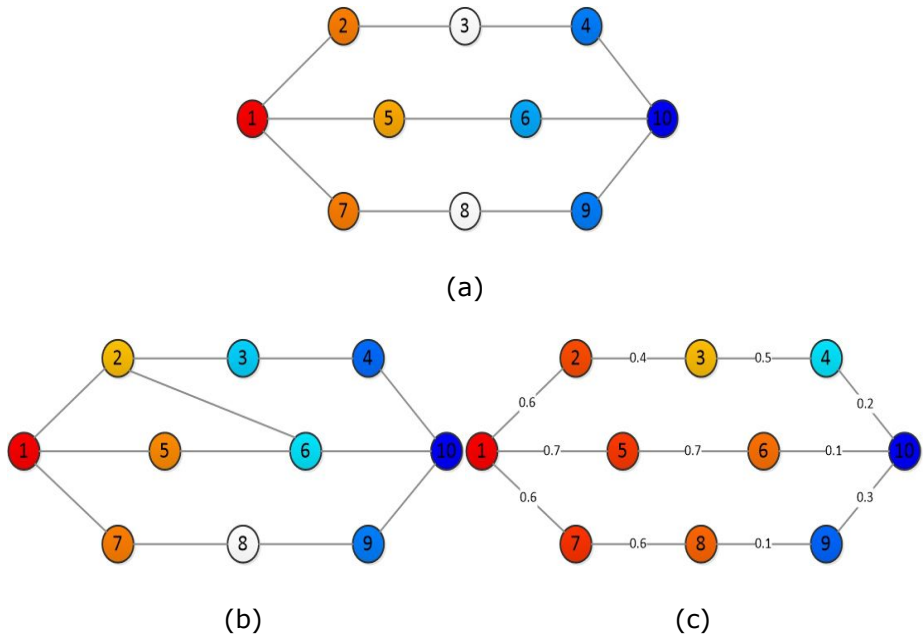

**Figure 1 | An example of how network structure influences the competition result.** (a) A simple undirected network of 10 agents with each edge of unit weight. The competition between agent 1 and agent 10 ends up as draw. (b) The network is derived from (a) by adding one edge between agent 2 and 6, which results in agent 10 being the winner. (c) The network has the same structure as network (a) but with different edge weights, which leads to agent 1 being the winner.

For network (a), $\Phi_{1,10}=0$, hence the competition ends up as draw. Network (b) is derived from network (a) by just adding one edge between agents 2 and 6, which results in $\Phi_{1,10}=-1$ and agent 10 being the winner. By changing weights of edges in network (a), we get network (c), which leads to $\Phi_{1,10}=3$ and agent 1 winning the competition. We can see that both network structure and coupling weights influence the competition results. In the following, we will focus on unweighted networks in the sense that the weight of every link in a network is one.

**Verification on a real network.** To see whether Eqs. (1)-(2) could properly model competition in real social networks, we test it on a commonly used benchmark model in social network analysis---the Zachary's karate club network[26] as shown in Fig. 2(a), which is a network of friendships between 34 members of a karate club at a US university in the 1970s. Due to the confliction between the manager (agent 34) and the coach (agent 1), the club finally splits into two communities, centered at the manager and the coach, respectively, as depicted by the vertical dashed line in Fig. 2(a).

In simulation, we fix the states of agents 1 and agent 34 at +1 and -1, respectively. The state of every other agent evolves according to Eq. (2). Fig 2(b) shows the steady states of all agents in the network, in which red agents are supporters of agent 1 and blue agents are supporters of agent 34. It is surprising to note that this splitting result completely matches the real situation as shown in Fig. 2(a). Furthermore, Fig. 2(b) also reveals the degree of supporting of each normal agent, represented by the darkness of the color. For example, agent 9 has the smallest absolute value of steady state among those supporters of agent 34, which implies that agent 9 is the weakest supporter of agent 34. This is also consistent with the reality that individual 9 is indeed the weakest political supporter of the manager[26]. Therefore, although our model is a very simplified version of the very complex real-world competition, it might be a reasonable mechanism for the competitive dynamics in some real social networks. Note that many network community detection methods can correctly reveal the two communities in the karate network[27], however, they do not explicitly use the information of the two competitors in the network and cannot reveal the degree of supporting of each agent towards the corresponding competitor.

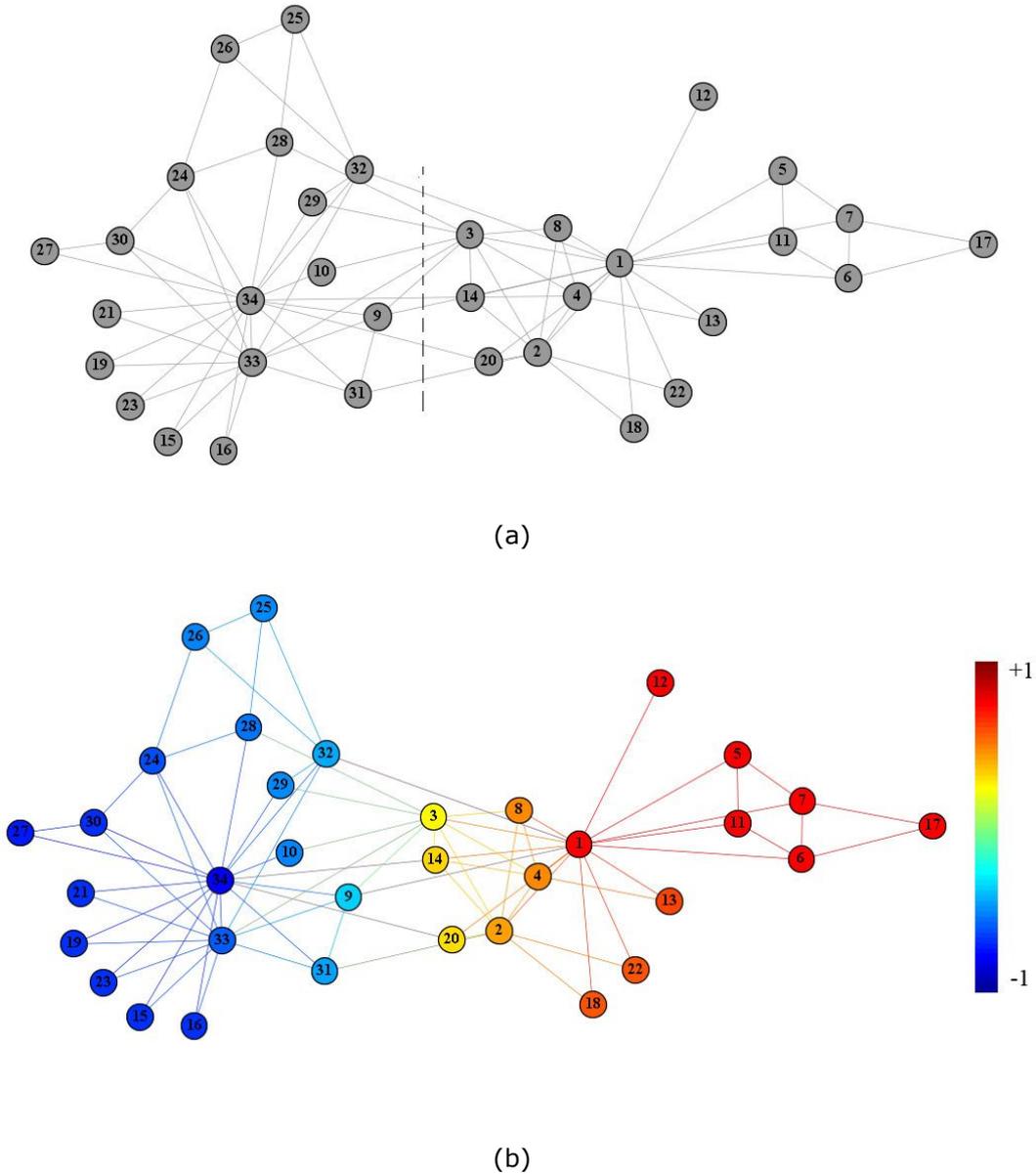

(a)

(b)

**Figure 2 | Verification of the model on Zachary's karate club network.** (a) Two real communities in the network led by agent 1 and agent 34, respectively, as divided by the dashed line in the figure. (b) Two communities derived from our model. Red community consists of supporters of agent 1 and blue community consists of supporters of agent 34. Darkness of the color represents the degree of supporting.

**Influence Matrix Criterion.** From the steady states expression in Eq. (3), competition results are fully determined by network structure and positions of the competitors in the network. However, directly computing the steady states according to Eq. (3) is computational inefficient for large-scale networks, since for every different pair of competitors, we have to re-compute the steady states. In the following, we compute the Influence Matrix (IM), in which each element characterizes the impact of one agent on another. Note that if there is a link from agent $k$ to $l$, i.e., $a_{kl}=1$, then agent $l$ has a direct impact on agent $k$. If there is a link from agents $k$ to $m$, and a link from agent $m$ to $l$, then agent $l$ has an indirect impact on agent $k$ via agent $m$. Intuitively, such an indirect impact should be weaker than the direct impact. Taking into account the fact that the number of paths of length $r$ from agent $k$ to $l$ is $(A^r)_{kl}$ in the unweighted network case, we define IM as a sum of the exponentially decreasing impact of increasingly paths:

$$F = I + \eta A + \eta^2 A^2 + \cdots \tag{5}$$

where $\eta \in (0, 1)$ is an attenuation factor. If $\eta \in (0, \lambda_1^{-1})$, where $\lambda_1$ is the largest eigenvalue of matrix $A$, then the above series converges[28] and we have:

$$F = (I - \eta A)^{-1}. \tag{6}$$

Let $f_{ki}$ be the entry of $F$ on $k$ th row and $i$ th column. Denote

$$\Gamma_{ij} \triangleq \sum_{k \in V/\{i,j\}} sgn(f_{ki} - f_{kj}), \tag{7}$$

We have the following IM criterion:

- **Which competitor will a normal agent support:** If $f_{ki} > f_{kj}$ ($f_{ki} < f_{kj}$), then agent $k$ will support competitor $i$ ($j$); If $f_{ki} = f_{kj}$, then agent $k$ is a neutral agent;
- **Which competitor will win:** If $\Gamma_{ij} > 0$ ($\Gamma_{ij} < 0$), then competitor $i$ ($j$) will win; If $\Gamma_{ij} = 0$, the competition ends up with a draw.

Although different choice of $\eta$ in Eq. (7) may generally result in different IM, we find that the IM criterion is robust with respect to $\eta$, in the sense that the criterion gives similar qualitative prediction for different choice of $\eta \in [0.5\lambda_1^{-1}, 0.9\lambda_1^{-1})$ (see Supplementary Figure S1). In the following simulations, we set $\eta = 0.85\lambda_1^{-1}$.

**Who will you support, and who will win from IM criterion.** For the Zachary's karate club network, Fig. 3 shows the difference $f_{k,1} - f_{k,34}$ between the influences of two competitors (agent 1 and agent 34) on a normal agent $k$. Comparing Fig. 3 with Fig. 2(b), we can see that $f_{k,1} - f_{k,34} > 0$ ( $f_{k,1} - f_{k,34} < 0$ ) if and only if $\bar{x}_k > 0$ ( $\bar{x}_k < 0$ ), which implies that the competition result can be fully predicted by the IM criterion in this case.

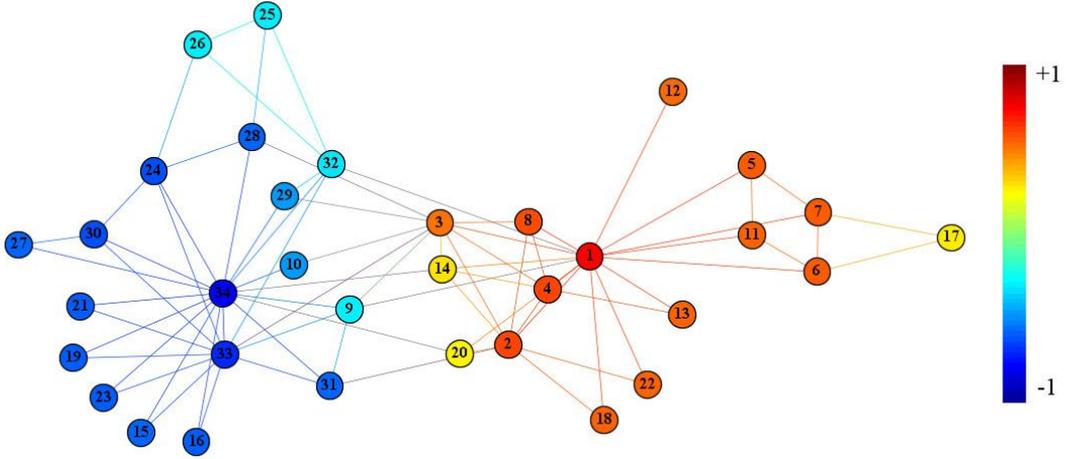

**Figure 3 | Application of the IM criterion to Zachary's karate club network.** Agent 1 and agent 34 are two competitors. A normal agent is colored red (blue) if the influence difference $f_{k1} - f_{k34} > 0$ ( $f_{k1} - f_{k34} < 0$ ). We dye all the nodes according to their normalized difference. The darker the color the larger the absolute difference is.

In general, for a given pair of competitors $i$ and $j$ in a network, we use the IM criterion to predict the bias of each normal agent and calculate the success rate of prediction as follows:

$$\rho_{ij} \triangleq \frac{1}{(N-2)} \sum_{k \in V/(i,j)} \frac{|sgn(f_{ik} - f_{jk}) + sgn(\bar{x}_k)||sgn(f_{ik} - f_{jk})||sgn(\bar{x}_k)| + g(|sgn(f_{ik} - f_{jk})| + |sgn(\bar{x}_k)|)}{2}, \quad (8)$$

where $g(x) = 2$, if $x = 0$; otherwise $g(x) = 0$. The average success rate of prediction on the bias of normal agents over all the $N(N-1)/2$ possible pairs of competitors in a network is denoted as $<\rho>$. Similarly, the success rate of prediction on who will win as the fraction of correct prediction over all the possible pairs of competitors can be formulated as follows:

$$\sigma \triangleq \frac{1}{N(N-1)} \sum_{i,j} \left| sgn(\Gamma_{ij}) + sgn(\Phi_{ij}) \right| \left| sgn(\Gamma_{ij}) \right| \left| sgn(\Phi_{ij}) \right| + g\left( \left| sgn(\Gamma_{ij}) \right| + \left| sgn(\Phi_{ij}) \right| \right). \quad (9)$$

Table I shows the value of $<\rho>$ and $\sigma$ for 15 real social networks. The maximum value of $<\rho>$ is 91.6%, the minimum is 74.0% and the average is 83.6%. $\sigma$ is almost always larger than 80%: the maximum is 96.9%, the minimum is 79.9% and the average is 86.0%. These results verify the validity of the IM criterion. We conjecture through simulation that for most pairs of competitors the prediction of a normal agent's bias being incorrect is because two competitors have very similar influence on the normal agent (see Supplementary Figure S2).

**Comparison with centrality-based criteria.** Given a pair of competitors, we can predict which competitor will win by the IM criterion. Intuitively, the winner should be more important or have higher impact on the network than the loser. Over the years, a number of centrality measures have been proposed to characterize the "importance" or "impact" of a node in a network. However, one difficulty in applying these centrality measures is that it is often unclear which of the many measures should be used in a particular circumstance. Here, we compare the IM criterion with criteria based on several common-used node centrality measures, including betweeness (BC), closeness (CC), degree (DC), eigenvector (EC), Katz (KC), K-Shell (KS) and PageRank (PR) (see Methods for the computation of these measures).

**Centrality-based criterion:** The competitor with higher centrality value will win. Competitors with the same centrality value will end up with a draw.

For each criterion, we calculate the success rate of prediction as the fraction of correct prediction of who will win over all $N(N-1)/2$ possible pairs of competitors. Fig. 4 and Fig. 5 show the success rate of prediction for 8 real undirected networks and 7 real directed networks, respectively. According to the average success rate over undirected and directed networks, we have the following order:

- For undirected networks: KC (84.8%), IM (84.4%), EC(79.7%), PR (78.4%), DC (77.8%), BC (69.4%), KS (61.4%), CC (39.6%).
- For directed networks: PR (93.0%), KC (88.3%), IM (88.0%), EC(87.0%), DC (79.8%), BC (77.3%), KS (61.0%), CC (37.5%).

We can see that criteria based on KC, PR, IM and EC are always better than the criteria based on the other four centralities. For undirected networks, KC criterion has the best performance: It provides highest success rate of prediction in 5 of 8 networks. On the other hand, PR criterion is always the best for each of the 7 directed networks. From the definition of KC, PR and EC, these results imply that whether a competitor could win depends to a large extent on both the number and importance of those agents that the competitor could directly influence.

In fact, the KC of node $i$ can be directly defined from IM as the influence of node $i$ on the whole network:

$$KC_i = \sum_{k \in V} f_{ki}. \qquad (10)$$

The KC-based prediction criterion can be derived from the IM criterion by just changing the order of summation and sign function in Eq. (7):

$$sgn(KC_i - KC_j) \triangleq sgn(KC_{ij}) = sgn \sum_{k \in V/\{i,j\}} (f_{ki} - f_{kj}), \qquad (11)$$

where $KC_i$ is the KC value of node $i$ (For a directed network, we just need to add one more term $(f_{ji} - f_{ij})$ in the sum). Directly summing up the influence errors in Eq. (11)

may help reduce perturbation, and thus result in more robust criterion. This might explanation why KC criterion is better to predict the winner than the IM criterion. PageRank is basically a variant of Katz centrality which is widely used for ranking nodes in directed networks such as WWW[29]. Although IM criterion is not the best, an advantage of IM criterion over node-centrality based criteria is that it could also predict the bias of each normal agent, in addition to predict the winner.

Degree (DC) is certainly the simplest criterion to predict the winner. However, it is a bit surprising to see that DC criterion provides as high as 80% success rate of prediction and performs even better than criteria based on BC, KS and CC. This implies that the number of agents that competitors could directly influence is still a relatively important factor. On the other hand, CC turns out to be the poorest criterion to predict the winner: the corresponding average success rate is just a little bit better than that of the completely random guessing (33.3%). Note that CC of a node captures how long it will take to spread information from the node to all other nodes sequentially. Our results show that this score has little effect on the competition.

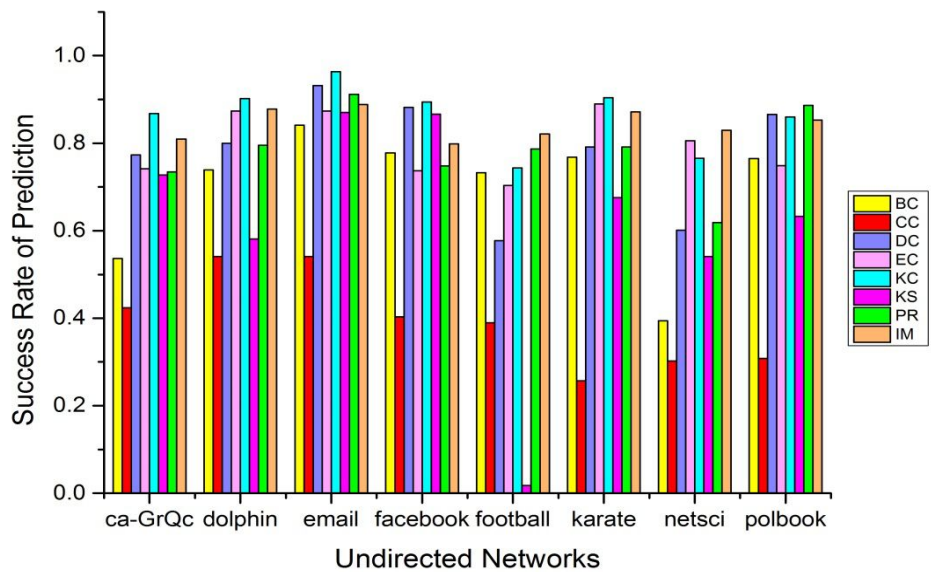

(a)

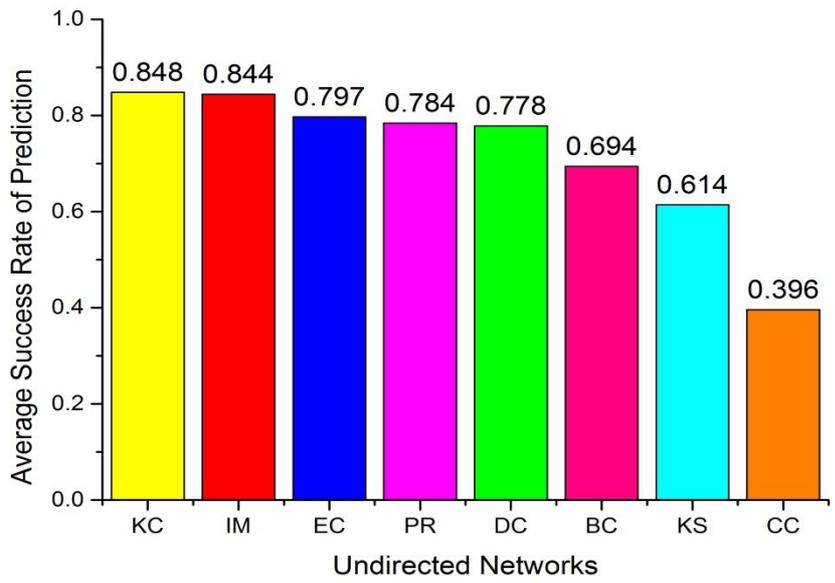

(b)

**Figure 4 | The success rate of prediction of competition result on 8 real undirected networks.** Here we compare the IM criterion with 7 centrality-based criteria. (a) the success rate of prediction for each network. (b) the average success rate of prediction of each criterion over 8 networks.

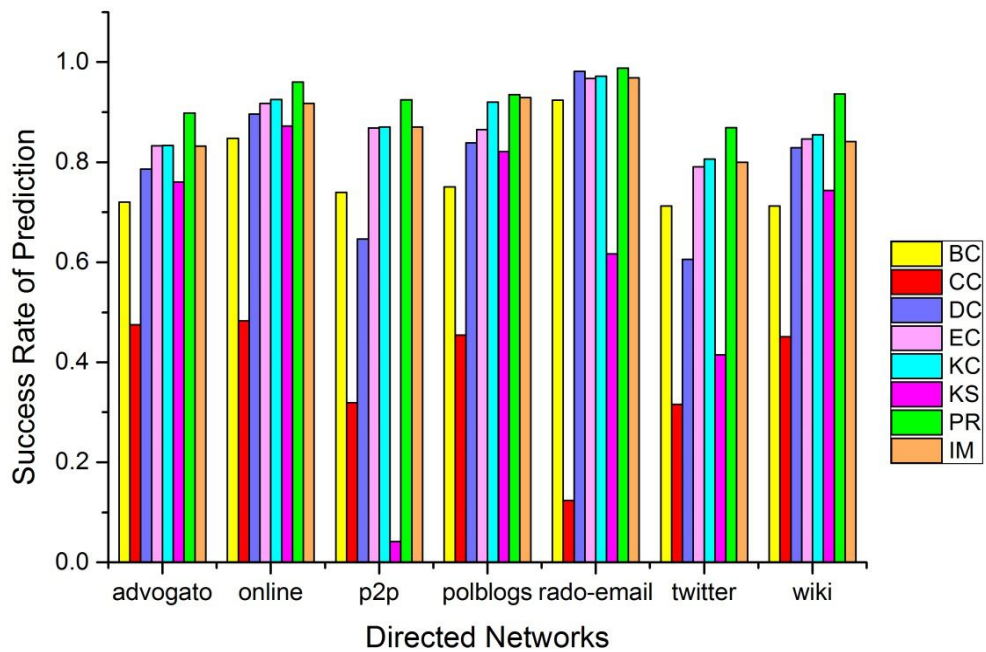

(a)

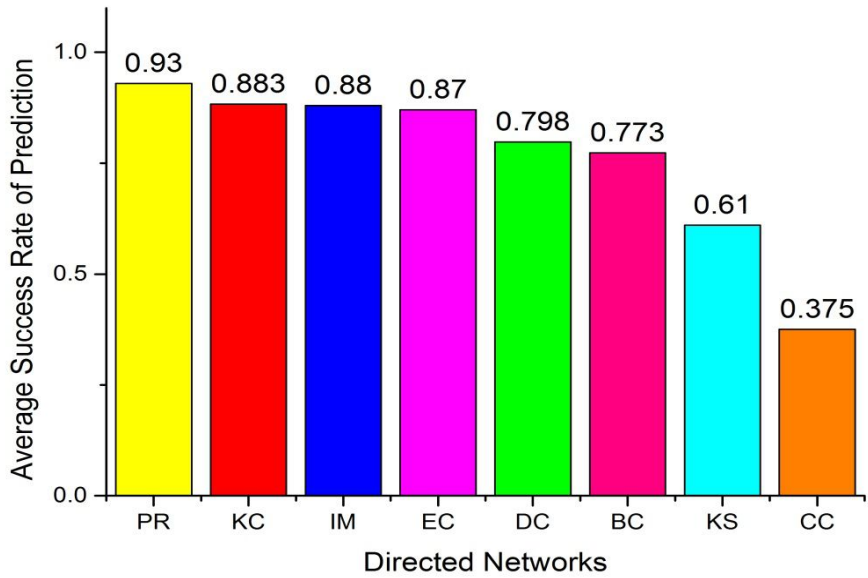

(b)

**Figure 5 | The success rate of prediction of competition result on 7 real directed networks.** Here we compare the IM criterion with 7 centrality-based criteria. (a) the success rate of prediction for each network. (b) the average success rate of prediction of each criterion over 7 networks.

**Discussion**

In summary, we study a model of competitive dynamics in which two competitors have fixed and different states, and each normal agent adjusts its state according to a distributed consensus protocol. The steady states of normal agents are fully determined by the network structure and positions of competitors in the network. Although real world competition involves a number of complex factors, we find that this very simple model can completely reveals the competition result in the well-known Zachary's karate club network. We investigate the Influence Matrix (IM) criterion to predict which competitor a normal agent will support and which competitor will win. We further compare the IM criterion with seven well-known node centrality measures. We find that Katz centrality (KC) and PageRank (PR) provide best prediction for undirected and directed networks, respectively.

These findings suggest that competitors in a network might use techniques such as PageRank optimization[30] to adjust network structure in order to win the competition. Although we assume that there are only two competitors in the model, the above analysis can also be generalized to the case with two sets of competitors in a network, and a nature way to deal with this case is to view all agents in a set as a super-agent. However, a key challenge here is that there does not existing a simple relationship between the sum of the centrality values of all agents in a set in the original network and the centrality score of the super-agent in the new network. All these issues will be considered in future work.

**Methods**

**Theoretical analysis of the model.** Eqs. (1)-(2) can be can be rewritten in the following matrix form:

$$X(t+1) = (I_N - \varepsilon H \times L)X(t) = TX(t),  \quad (12)$$

where $I_N$ is an identity matrix; $L = D - A$ is the Laplacian matrix, $D$ is the diagonal matrix of agents' out-degrees; $H$ is an indicative diagonal matrix with $H(s,s) = 0$ if agent $s$ is a competitor, and $H(s,s) = 1$ otherwise. Obviously, the sum of each row of matrix $T$ equals to 1.

For convenience, we reorder the agents so that the two competitors come last. Thus, we have

$$D = \begin{bmatrix} \bar{D} & 0 & 0 \\ 0 & d_i & 0 \\ 0 & 0 & d_j \end{bmatrix} \text{ and } A = \begin{bmatrix} \bar{A} & \mathbf{c}_i & \mathbf{c}_j \\ \mathbf{r}_i & 0 & * \\ \mathbf{r}_j & * & 0 \end{bmatrix}, \quad (13)$$

where $d_i$ and $d_j$ denote the out-degrees of competitor $i$ and $j$, respectively; vectors $\mathbf{c}_i$, $\mathbf{c}_j$, $\mathbf{r}_i$, and $\mathbf{r}_j$ contain the corresponding elements in the reordered coupling matrix.

Hence, Eq.(12) can be rewritten as

$$\begin{bmatrix} X_{norm}(t+1) \\ x_i(t+1) \\ x_j(t+1) \end{bmatrix} = \begin{bmatrix} Q & B \\ 0 & 1 & 0 \\ 0 & 0 & 1 \end{bmatrix} \begin{bmatrix} X_{norm}(t) \\ x_i(t) \\ x_j(t) \end{bmatrix}, \quad (14)$$

where $X_{norm} \in R^{N-2}$ represents the state vector of all normal agents; $Q = I_{N-2} - \varepsilon(\bar{D} - \bar{A})$ and $B = \varepsilon \begin{bmatrix} \mathbf{c}_i & \mathbf{c}_j \end{bmatrix}$. Thus,

$$X_{norm}(t) = QX_{norm}(t-1) + B\begin{bmatrix} x_i(t-1) \\ x_j(t-1) \end{bmatrix}$$
$$= Q^t X_{norm}(0) + \sum_{k=0}^{t-1} Q^k B\begin{bmatrix} x_i(0) \\ x_j(0) \end{bmatrix} \qquad (15)$$

If each normal agent has a path connecting to at least one competitor, then $(\bar{D}-\bar{A}) \in R^{N-2}$ is invertible[31]. Since $0 < \varepsilon < D_{max}^{-1}$, we can show from Geršgorin disk theorem that the spectral radius of $Q$ is less than 1. Thus, as $t \to \infty$, we have

$$X_{norm}(t) \to (I_{N-2}-Q)^{-1} B \begin{bmatrix} x_i(0) \\ x_j(0) \end{bmatrix}$$
$$= (I_{N-2}-I_{N-2}+\varepsilon\bar{D}-\varepsilon\bar{A})^{-1}\varepsilon \begin{bmatrix} \mathbf{c}_i & \mathbf{c}_j \end{bmatrix} \begin{bmatrix} x_i(0) \\ x_j(0) \end{bmatrix} \qquad (16)$$
$$= (\bar{D}-\bar{A})^{-1} \begin{bmatrix} \mathbf{c}_i & \mathbf{c}_j \end{bmatrix} \begin{bmatrix} +1 \\ -1 \end{bmatrix}.$$

According to Lemma 4 in Ref. 32, each entry of $(\bar{D}-\bar{A})^{-1}\begin{bmatrix} \mathbf{c}_i & \mathbf{c}_j \end{bmatrix}$ is nonnegative and each row sum of $(\bar{D}-\bar{A})^{-1}\begin{bmatrix} \mathbf{c}_i & \mathbf{c}_j \end{bmatrix}$ is equal to one. Thus, the steady state of each normal agent is a convex combination of +1 and -1.

**Computation of centrality measures.** As in our definition of the network structure, an link from agent $k$ to agent $l$ means agent $k$ could be directly influenced by agent $l$. Hence, for directed networks, we compute the centrality measures as follows (which can be also applied to undirected networks):

In-BC[33]: the in-betweenness centrality of node $l$ is computed by $CB_{in}(l) = \sum_{k<m} \frac{g_{km}(l)}{g_{km}}$, where $g_{km}$ is the number of geodesics from node $k$ to node $m$ and $g_{km}(l)$ is the number of geodesics that node $l$ is on;

In-CC[34]: the in-closeness centrality of node $l$ is computed by $CC_{in}(l) = \sum_{k \neq l} \frac{1}{d(k,l)}$, where $d(k,l)$ is the shortest distance from node $k$ to node $l$;

In-DC: the in-degree of a node is the number of agents that an agent could directly influence;

In-EC[21]: the eigenvector centrality is a natural extension of degree by considering both the number and the importance of those agents that an agent could directly influence. The EC of a network is equal to the eigenvector corresponding to the largest eigenvalue of the coupling matrix. According to the definition of the network structure, we use $A^T$ to compute the In-EC;

In-KC: the Katz Centrality is a variation of EC, by adding an initial importance to each agent. The In-Katz-Centrality of a network is computed by $(I-\alpha A^T)^{-1}\mathbf{1}$, where $\mathbf{1}$ is a vector with all ones of an appropriate size, and the attenuation factor. Related studie[35] shows that there is no significant change in ranking of nodes based on Katz Centrality with $\alpha \in [0.5\lambda_1^{-1}, \ 0.9\lambda_1^{-1})$. In simulations, we set $\alpha = 0.85\lambda_1^{-1}$;

In-KS[36]: nodes are assigned to different in-shells according to their remaining in-degrees, which is obtained by successive pruning of nodes with in-degree smaller than the current in-k-shell value. We start by removing all nodes with in-degree $k_{in} \leq 1$, until that all nodes left are with in-degree larger than 1. The removed nodes, along with the corresponding links, form an in-k-shell with index $k_{ins} = 1$. In a similar fashion, we iteratively remove the next in-k-shell. As a result, each node is associated with one $k_{ins}$ index;

PageRank: the algebraic expression of the page rank can be formulated as $PR = (I - \mu A^T D^{-1})^{-1} \mathbf{1} \frac{1-\mu}{N}$, where $\mu$ is the dampening factor. We use the power method[37] to compute the page rank value, and set $\mu = 0.85$. The Page Rank is a variation on the Katz Centrality by dividing the importance of those agents which could directly influenced by an agent, by their out-degrees.

**Acknowledgments**

This work was supported by the National Natural Science Foundation of China under Grant Nos. 61374176 and 61104137, the Science Fund for Creative Research Groups of the National Natural Science Foundation of China (No. 61221003), and the National Key Basic Research Program (973 Program) of China (No. 2010CB731403).

**Author contribution**

X.W. envisioned the study. J.Z., Q.L. and X. W. conceived the theoretical analysis. J.Z. designed the experiments and performed the computational analysis. J.Z., Q.L. and X.W. wrote the manuscript.

**Additional information.**

**Competing financial interests:** The authors declare no competing financial interests.

**Figure Legends**

**Figure 1 | An example of how network structure influences the competition result.** (a) A simple undirected network of 10 agents with each edge of unit weight. The competition between agent 1 and agent 10 ends up as draw. (b) The network is derived from (a) by adding one edge between agent 2 and 6, which results in agent 10 being the winner. (c) The network has the same structure as network (a) but with different edge weights, which leads to agent 1 being the winner.

**Figure 2 | Verification of the model on Zachary's karate club network.** (a) Two real communities in the network led by agent 1 and agent 34, respectively, as divided by the dashed line in the figure. (b) Two communities derived from our model. Red community consists of supporters of agent 1 and blue community consists of supporters of agent 34. Darkness of the color represents the degree of supporting.

**Figure 3 | Application of the IM criterion to Zachary's karate club network.** Agent 1 and agent 34 are two competitors. A normal agent is colored red (blue) if the influence difference $f_{k1} - f_{k34} > 0$ ($f_{k1} - f_{k34} < 0$). We dye all the nodes according to their normalized difference. The darker the color the larger the absolute difference is.

**Figure 4 | The success rate of prediction of competition result on 8 real undirected networks.** Here we compare the IM criterion with 7 centrality-based criteria. (a) the success rate of prediction for each network. (b) the average success rate of prediction of each criterion over 8 networks.

**Figure 5 | The success rate of prediction of competition result on 7 real directed networks.** Here we compare the IM criterion with 7 centrality-based criteria. (a) the success rate of prediction for each network. (b) the average success rate of prediction of each criterion over 7 networks.

**Table I | The average success rate of prediction of the IM criterion on 15 real networks.** For each network, we show its type and name; number of nodes (N) and links (M) of the largest strongly connected component; the average success rate of prediction on the bias of normal agents ($<\rho>$) and the success rate of prediction on who will win ($\sigma$).

| Type | Name | N | M | $<\rho>$ | $\sigma$ |
|---|---|---|---|---|---|
| Undirected | ca-GrQc[38] | 4158 | 13428 | 0.740 | 0.809 |
| Undirected | dolphin[39] | 62 | 159 | 0.878 | 0.878 |
| Undirected | email[40] | 1133 | 10902 | 0.847 | 0.888 |
| Undirected | facebook[41] | 4039 | 88234 | 0.754 | 0.799 |
| Undirected | football[42] | 115 | 4120 | 0.834 | 0.821 |
| Undirected | karate[26] | 34 | 78 | 0.902 | 0.872 |
| Undirected | netsci[43] | 379 | 1828 | 0.842 | 0.829 |
| Undirected | polbook[44] | 105 | 441 | 0.821 | 0.853 |
| Directed | advogato[45] | 3140 | 40066 | 0.818 | 0.832 |
| Directed | online[46] | 1294 | 19026 | 0.897 | 0.917 |
| Directed | p2p[47] | 2068 | 9313 | 0.847 | 0.870 |
| Directed | polblogs[48] | 793 | 15781 | 0.853 | 0.929 |
| Directed | rado-email[49] | 126 | 5639 | 0.916 | 0.969 |
| Directed | twitter[50] | 1726 | 6901 | 0.781 | 0.800 |
| Directed | wiki[51] | 1300 | 39456 | 0.816 | 0.841 |